\documentclass[12pt]{article}
\pdfoutput=1
\usepackage{graphicx}
\usepackage{epstopdf}
\usepackage{amsmath}
\usepackage{amsfonts}
\usepackage{amssymb}
\usepackage{color}
\usepackage{mathrsfs}

\newcommand{\be}{\begin{equation}}
\newcommand{\ee}{\end{equation}}
\newcommand{\bea}{\begin{eqnarray}}
\newcommand{\eea}{\end{eqnarray}}
\newcommand{\barr}{\begin{array}}
\newcommand{\earr}{\end{array}}

\def\beq{\begin{equation}}
\def\eeq{\end{equation}}
\def\be{\begin{equation}}
\def\ee{\end{equation}}
\def\bea{\begin{eqnarray}}
\def\eea{\end{eqnarray}}

\def\mpl{M_{\rm Pl}}

\begin{document}

%\begin{titlepage}

\setcounter{page}{1} \baselineskip=15.5pt \thispagestyle{empty}

\begin{flushright}
%hep-th/yymmnnn\\
\end{flushright}
%\vfil

\begin{center}

{\Large \bf On Loops in Inflation II:\\[0.3cm] IR Effects in Single Clock Inflation}
\\[0.7cm]
{\large Leonardo Senatore${}^{1,2}$  and Matias Zaldarriaga${}^3$}
\\[0.7cm]
%\vspace{.7cm}
%\vspace{.3cm}
{\normalsize { \sl $^{1}$ Stanford Institute for Theoretical Physics, Stanford University, Stanford, CA 94306}}\\
\vspace{.3cm}

{\normalsize { \sl $^{2}$ Kavli Institute for Particle Astrophysics and Cosmology, Stanford University and SLAC,\\ Menlo Park, CA 94025}}\\
\vspace{.3cm}

{\normalsize { \sl $^{3}$ School of Natural Sciences, Institute for Advanced Study, \\Olden Lane, 
Princeton, NJ 08540, USA}}\\
\vspace{.3cm}

\end{center}

\vspace{.8cm}

\hrule \vspace{0.3cm}
{\small  \noindent \textbf{Abstract} \\[0.3cm]
\noindent
In single clock models of inflation the coupling between modes of very different scales does not have any significant dynamical effect during inflation. It leads to interesting projection effects.  Larger and smaller modes change the relation between the scale a mode of interest will appear in the post-inflationary universe and will also change the time of horizon crossing of that mode. We argue that there are no infrared projection effects in physical questions, that there are no effects from modes of longer wavelength than the one of interest at the time of reheating. These potential effects cancel when computing fluctuations as a function of physically measurable scales. Modes on scales smaller than the one of interest change the mapping between horizon crossing time and scale. The correction to the mapping computed in the absence of fluctuations is enhanced by a factor $N_e$, the number of $e$-folds of inflation between horizon crossing and reheating.  The new mapping is stochastic in nature but its variance is not enhanced by $N_e$.
 \vspace{0.3cm}
\hrule
%\vfil
%\begin{flushleft}
%\today
%March 20, 2008
%\end{flushleft}

%\end{titlepage}

%\newpage
%\tableofcontents
%\newpage

\section{Introduction}

Loop corrections to inflationary correlators can have interesting infrared (IR) behaviors. In perturbation theory, depending on the kind of interactions considered, it is possible to find logarithmic running of the form $\log(H/\mu)$~\cite{Senatore:2009cf}, where $\mu$ is the renormalization scale and $H$ is the Hubble scale and represents the natural low energy scale within the logarithmic corrections. One also finds time-dependent corrections~\cite{Weinberg:2005vy,Weinberg:2006ac, Kahya:2010xh}, corrections that grow with the time of observation. And finally there are also  corrections proportional to $\log(k L)$, where $k$ is the comoving wavenumber of the mode considered, while $L$ is a comoving infrared scale~\cite{Giddings:2010nc,Byrnes:2010yc}. The purpose of this paper is to address these last infrared terms in the context of single clock models of inflation. 

We will show that they fall into two classes. The first class of infrared divergencies results from asking an unphysical question and disappear after a redefinition of the observable quantities. The second class has never been identified before, and they are actually physical, in the sense that they lead to measurable effects. They are projection type effects caused by modes that are shorter wavelength than the one of interest and change the mapping between observed scale and time of horizon crossing. These new corrections we identify are enhanced by a factor of $N_e$, the number of $e$-folds between the point in the inflationary history a mode crosses the horizon and the end of inflation.  Although this new  mapping is stochastic, its variance is small, not enhanced by $N_e$ and can thus be thought as deterministic. 

The effect we uncover is just the perturbative version of the same physics that leads to slow-roll eternal inflation. 
We will first find an expression for the projection effect caused by the short modes at lowest order in the amplitude of fluctuations. However this relation to slow-roll eternal inflation will allow us to borrow from that literature and re-sum the expressions to obtain a result valid even in cases where the product of the amplitude of fluctuations times $N_e$ is large. 

This paper contains results presented in a talk at the Perimeter Institute in June 2010 which is available online at http://pirsa.org/10100079/. The results have already been cited in the literature ({\it e.g.}~\cite{Gerstenlauer:2011ti,Giddings:2011zd, Giddings:2011ze}).  For clarity, additional results we obtained after that talk can be found in our companion paper [10]. After our talk at the Perimeter Institute, other papers on topics related to the ones treated here have appeared in the literature. For example references~\cite{Gerstenlauer:2011ti,Giddings:2011zd, Giddings:2011ze,Chialva:2011bg,Tanaka:2011aj,Urakawa:2010kr,Urakawa:2011fg,Riotto:2011sf} discuss the effect on the power spectrum of modes longer than the one observed. Reference~\cite{Riotto:2011sf} discusses some of these issues using the stochastic approach to inflationary perturbations. Instead, references~\cite{Giddings:2011zd, Giddings:2011ze} discuss the effect on the power spectrum of modes shorter than the one observed. Since in this paper we discuss only the results presented in the talk  at the Perimeter Institute, we do not describe the subsequent literature in further detail, but we refer the interested reader to read it directly.

\section{Dynamical effects\label{sec:dynamical}}

We are interested in understanding how the presence of $\zeta$ fluctuations at a given scale affects the behavior of $\zeta$ correlators on a different scale. 
In this section we will consider potential dynamical effects in the context of single field inflationary models. For example, dynamical effects might include a time dependence of the $\zeta$ correlators on super horizon scales, or a change in the amplitude of correlations at horizon crossing due to the interaction with other modes. We distinguish these dynamical effects from projection effects we will consider in the next section. We call projection effects simply changes in the scale at which an observer in the late universe will measure a particular mode while dynamical effects are at least in principle observable during inflation~\footnote{For example, one could consider the amplitude of modes just before Horizon crossing as a function of the proper time of a free falling observer or the expansion rate for a given value of the scalar field as quantities that would be changed by what we call a dynamical effect during inflation and could in principle be observed at the time. }. We make this distinction because dynamical effects are basically zero and furthermore projection effects are a bit less well defined as they depend on the properties of the Universe after inflation.

To organize our discussion we will consider in turn the potential dynamical effects from modes longer and shorter than the one of interest. 

\paragraph{Shorter wavelength modes:}

As proven in \cite{loops1} the values of $\zeta$ correlators at a given scale become constant once that scale crosses the horizon. Shorter modes, that cross the horizon later than the mode of interest, do not have any dynamical effect on the mode once the background evolution is specified.  Short wavelength fluctuations generate  tadpoles that modify the background  solution. In this paper $\zeta$ describes fluctuations around a background history that satisfies the homogeneous equations of motion. Once we have modified the background history or added suitable counter terms to cancel the tadpoles as we did in \cite{loops1}, $\zeta$ correlators are time-independent once out of the horizon. 

\paragraph{Longer wavelength modes:}

$\zeta$ becomes constant in time once a mode has a wavelength  longer that the horizon. We will call these modes background modes, $\zeta_B$, as locally they can be reabsorbed in a redefinition of the quantities in the background solution. If one is interested in the dynamics of shorter wavelength modes in this background one can simply include the effect of the background mode by replacing $a(t) \rightarrow e^{\zeta_B} a(t)$ in the action for the short modes. The corrections to this statement  are suppressed by at least one factor of $(k_L / a(t) H(t))^2$ where $k_L$ is the wavenumber of the background mode~\cite{Ganc:2010ff,Creminelli:2011rh}. 

In the presence of a long mode we can write the metric as:
 \be
ds^2=-dt^2+a(t)^2 e^{2\zeta_B}d\vec x^2\ ,
\ee 
which means that a background $\zeta_B$ mode amounts just to an overall rescaling of the spatial coordinates: 
\be\label{rescalex}
\Delta x \rightarrow e^{\zeta_B}\Delta x\ .
\ee 
Thus the correlation function on scales short compared to the wavelength of the background mode,  in the presence of that background mode $\xi_B(\Delta x)$ is directly related to the one in the absence of it, $\xi_0(\Delta x)$ by:
\be
\xi_B(\Delta x) \equiv \langle\zeta\zeta\rangle |_{\zeta_B}(\Delta x)=\langle\zeta\zeta\rangle|_{\zeta_B=0} (e^{-\zeta_B}\Delta x) \equiv \xi_0(e^{-\zeta_B} \Delta x)\ .
\ee
This formula is valid for all short distances, even distances inside the horizon. It is also valid during inflation.
 
The power spectrum is related the the correlation function by:
\be
P_B(k) = \int d^3x\, e^{ik x} \xi_B(x)\ , 
\ee
which implies that
\be\label{shiftDelta}
\Delta^2_B(k) \equiv {k^3 P_B(k) \over 2 \pi^2} = \Delta^2_0(k e^{-\zeta_B})\equiv{ (e^{-\zeta_B}k)^3 P_0(e^{-\zeta_B}k) \over 2 \pi^2}\ ,
\ee
where $P_0$ is the power spectrum in the absence of $\zeta_B$. 
It is important to note that this relation is valid in any given realization of the long modes. No average over the long modes has been taken. Notice also that $\Delta_0\sim 1$ close to the eternally inflating regime, when this rescaling takes its largest numerical value. 

Although the above formulas are valid for modes inside and outside the horizon and at any point during inflation as long as the background mode is super-horizon, we are interested here in the amplitude of fluctuations when they cross the horizon, when they freeze. We have already mentioned that there is no time evolution after that point.  At leading order in the slow roll parameters, the amplitude of fluctuations after horizon crossing in the absence of the background mode is simply given by:
\be
\Delta^2_0=  {1 \over 8 \pi^2} \left.\frac{H^4}{|\dot H| \mpl^2}\right|_{t_{hc}},
\ee
where the expression on the right hand side needs to be evaluated at the time $t_{hc}$ when the comoving mode $k$ crossed the horizon, 
$k=a(t_{hc}) H(t_{hc})$. For simplicity we are just quoting the formula valid for standard slow-roll inflationary models but our arguments are valid in general. The reader may replace this by the formula applicable in their favorite single clock inflationary model~\cite{Cheung:2007st}. 

Thus equation (\ref{shiftDelta}) implies that this expression remains valid in the presence of the background mode,
\be\label{amplitude_back}
\Delta^2_B =  {1 \over 8 \pi^2} \left.\frac{H^4}{|\dot H |\mpl^2}\right|_{t_{hc}},
\ee
as long as we compute the horizon crossing time properly: $k=a(t_{hc})e^{\zeta_B} H(t_{hc})$ as expected from eq.~(\ref{rescalex}). 

Equation (\ref{amplitude_back}) states that the amplitude of a mode only depends on the properties of the background at horizon crossing and it is independent of the value of the longer wavelength modes. If one specifies the moment of horizon crossing then there is no dependence on the longer modes even in a given realization. Infrared logarithms appear only when one computes correlation functions at a specified comoving momenta rather than for modes labelled by the time of horizon crossing~\footnote{Related considerations regarding this point have also appeared in~\cite{Byrnes:2010yc,Urakawa:2010kr}. We stress that there is no physical effect at the reheating surface from modes longer than the horizon in every single realization, not just on average. However, it should be made clear that our statement does not mean that calculations of the $\zeta$ power spectrum at reheating done using comoving momenta and that show an IR divergence~\cite{Giddings:2010nc,Byrnes:2010yc} are incorrect. As we discuss in the Appendix, they are correct. Indeed, one could compute a physical observable by using comoving coordinates and by evolving non-linearly the fluctuations up to the time of observation including all relevant projection type effects such as lensing, gravitational redshifts, etc. One could follow IR divergencies in intermediate results but in the end see that the final result has no IR divergencies.  The IR divergencies in the power spectrum at reheating cancel out with effects coming from the three- and four-point functions of $\zeta$ at reheating. Our way of presenting the results makes this manifest already in the intermediate results at reheating, proving in this way that the IR divergencies from modes longer than the horizon are not harmful. It is still true that in order to compute an observable quantity, one should evolve our result non-linearly up to the detector.'}. This is discussed further in the Appendix. Given that the amplitude of a particular mode is fixed at horizon crossing one needs to determine the observed scale for a mode which crossed the horizon at a given point during the evolution. We refer to this mapping between observed scale and horizon crossing time as projection which we discuss in the next section.

\section{Projection Effects\label{sec:kinematics}}

Since the amplitude of fluctuations at a given scale is only determined by what was happening at horizon crossing during inflation we need to find the mapping between horizon crossing time and scale measured by an observer in the post-inflationary universe. We refer to this mapping as projection. In detail this mapping might be complicated and depend on the experimental probe this late time observer uses. Depending on how the experiment is done, eminently late time effects such as gravitational lensing or redshift distortions would affect this mapping. Furthermore such a calculation will depend on the expansion history after inflation and the time at which the observation is made. 

Perhaps more importantly, these late time projection effects are well studied in the astrophysical literature, the magnitude has been computed and the most important ones such as gravitational lensing have been thoroughly studied and already detected in the CMB and other data. 
Thus we want to circumvent as much as possible the late time projection effects and calculate properties at the reheating surface. We will still calculate something that is directly related to what observers will measure. At least for some probes, like the CMB bispectrum, calculations in the literature already account for all the projection effects 
%{\bf ?!?!?!??!??!??!!?!? isn't the last sentence somewhat a disconnected repetition of two sentences above? Can we cut it or I am misundertanding ? ?!?!?!?!!?!?!?1/}. 

The fundamental problem is to relate the comoving coordinates used in the calculation to a more physical measure of length. To do so we will simply use the physical volume of a region. This procedure corresponds to counting particles present in the region after reheating and using the enclosed number of particles as a measure of the physical size of the region~\footnote{Perhaps it might be suggested that to avoid ambiguities we should count entropy which is conserved in the subsequent FRW evolution rather than particles. However this distinction has little relevance, all that is important is that the late time observer uses the number of particles of some species to establish a measure of scale. Any such measure is proportional to the volume of the region at reheating.}. Thus we are asking questions about the curvature of surfaces of constant temperature in the late universe as a function of their size as measured by the number of particles contained in the region. In the late universe we could use the number of dark matter particles or equivalently the mass in dark matter particles enclosed in the volume. Not only is this measure of size a well defined physical choice, but the amplitude of perturbations at a given mass scale is directly related to the abundance of objects of that mass that will form in the late Universe. 

Equivalently one could think of the case of the CMB and consider the amplitude of fluctuations at the first acoustic peak. That scale corresponds to the sound horizon at the epoch of recombination. This physical scale, the scale of the acoustic peak, contains a number of particles that can easily be calculated. For example the number of CMB photons is given by $N_{CMB} \propto T_{CMB}^3 c_S^3 H^{-3}$, where $c_S$ is the sound speed and all quantities are evaluated at recombination. So when computing the amplitude of fluctuations as a function of scale measured in terms of the acoustic scale, one is again basically using the enclosed number of particles to define length scales. The number of enclosed particles can then be easily mapped into a volume in the reheating surface~\footnote{Our comments about using entropy apply here but just as before this complication is immaterial.}. 

To find the mapping we first write down the range of comoving coordinates that correspond to a Hubble volume at the moment of horizon crossing,
\be
H(t_{hc})^{-3}= a^3(t_{hc}) e^{3 \zeta(t_{hc})} V_x\ ,
\ee
where $V_x$ is the comoving volume of the region and $\zeta$ has been taken to be constant across the region.  
Curvature fluctuations at this scale are given by 
\be\label{amplitude}
\Delta^2=  {1 \over 8 \pi^2} \left.\frac{H^4}{|\dot H |\mpl^2}\right|_{t_{hc}}\ .
\ee 
At reheating this region of comoving coordinates encloses a physical volume $V_{rh}$ given by
\be\label{reheating-volume}
V_{rh}= a^3(t_{rh})\int d^3x\; e^{3 \zeta(t_{rh})}  = {a^3(t_{rh}) \over a^3(t_{hc})}H(t_{hc})^{-3} \int {d^3 x \over V_x}\, e^{3 (\zeta(t_{rh})-\zeta(t_{hc}))}\ .
\ee
We note that $\zeta(t_{rh})$ is not constant across the volume but we have suppressed its $x$ dependence to keep the notation simple. 
This equation implies that fluctuations on scale larger than the one we are considering $(k V_x^{1/3} \ll 1)$, do not have any influence on the projection effects we are calculating because for those scales $\zeta$ takes the same value at horizon crossing and reheating and thus cancels in equation (\ref{reheating-volume}):
\be
\zeta(t_{rh})-\zeta(t_{hc})=\int \frac{d^3k}{(2\pi)^3}\,\left(\zeta_k(t_{rh})-\zeta_k(t_{hc})\right)\simeq\int_{k V_x^{1/3}\gtrsim 1} \frac{d^3k}{(2\pi)^3}\;\left(\zeta_k(t_{rh})-\zeta_k(t_{hc})\right)\ .
\ee

We emphasize again that this does not mean that this longer modes would not affect the measurements of a late time observer. Provided the modes are shorter than the horizon at the time of the observation they will contribute to the late time projection effects ({\it eg.} gravitational lensing) as calculated in the standard fashion. It is important to also note that once the modes are inside the horizon in the late Universe they also have dynamical effects on other modes which are also inside the horizon. Equation (\ref{reheating-volume}) just makes the perhaps trivial point that modes larger than the horizon in addition to not having any dynamical effects also do not have any projection effects. 

If one were to calculate the reheating volume of a fixed range of comoving scales there would be no cancelation and one would encounter an infrared divergence, but this divergence is just the result of not calculating a physically relevant quantity. 
The divergence comes from comparing power at widely different physical scales in different regions of the Universe. As the length of inflation increases, the variance in the relation between a comoving scale and a physical scale becomes progressively larger, leading to an IR divergence in the amplitude of fluctuations at fixed comoving scale. 

Modes on scales shorter that the one of interest contribute to the projection effects as they do not cancel in the difference $\zeta(t_{rh})-\zeta(t_{hc})$.
In equation (\ref{reheating-volume}) the reheating volume is a stochastic variable that changes from realization to realization~\footnote{Strictly speaking $V_{rh}$ is a quantum operator. However because we are dealing with modes that are larger than the horizon this distinction is not important. Furthermore although we have concentrated on situations where the fluctuations are quantum mechanical in nature, our conclusion also apply to examples where fluctuations are basically classical such those treated in \cite{Nacir:2011kk}.}. In order to get a first estimate of the effect we will first consider cases for which $\langle \zeta^2 \rangle \ll1$ so we can expand the exponential in equation (\ref{reheating-volume}) in a Taylor series. We can then write
\bea\label{reheating-volume2}
V_{rh}&=& {a^3(t_{rh}) \over a^3(t_{hc})}H(t_{hc})^{-3} \int {d^3 x \over V_x} \left(1 + 3 (\zeta(t_{rh})-\zeta(t_{hc}) + {9\over 2}(\zeta(t_{rh})-\zeta(t_{hc})^2+ \cdots\right) \nonumber \\ 
&\approx& {a^3(t_{rh}) \over a^3(t_{hc})}H(t_{hc})^{-3}  \left(1 + {9\over 2}  {1\over V_x} \int_{k V_x^{1/3} > 1}  {d^3k \over (2 \pi)^3} |\zeta(k)|^2 + \cdots \right)\ .
\eea
We explicitly show that only short modes contribute to the integral and we have used that for this modes the spatial integral of $e^{i(k-k^\prime) x}$ over the volume can be approximated by a delta function. Equivalently we could have expanded the short wavelength fluctuations inside of $V_x$ using a orthonormal set of modes over the region, or assume periodic boundary conditions. The results are independent of these choices. 

Though this is not strictly necessary, to make further progress let us take the expectation value of  (\ref{reheating-volume2}). This will turn out to capture the leading result as the variance of this quantity is usually negligible. We get: 
\be\label{eq:kinematical-one-average}
 \langle V_{rh} \rangle \approx {1 \over H(t_{hc})^3 } {a^3(t_{rh}) \over a^3(t_{hc})}  \left(1 + {9\over 2}  {1\over V_x} \int_{k V_x^{1/3} > 1}  {d^3k \over (2 \pi)^3} \langle |\zeta(k)|^2 \rangle+ \cdots \right)\ .
\ee
We can now use that
\be
\frac{a(t_{\rm rh})}{a(t)}=e^{N_e}\ ,
\ee
where $N_e$ is the classical number of $e$-foldings from time $t$ to the reheating time $t_{rh}$. For an approximately scale invariant spectrum we also have that 
\be
{1\over V_x} \int_{k'^3 V_x > 1}  {d^3k' \over (2 \pi)^3} \langle |\zeta(k)|^2 \rangle = \int_{k^\prime > k} d\ln k^\prime\; \Delta^2 \approx \Delta^2 N_e \ ,
\ee
so that
\be\label{eq:kinematical-one-taylor-Ncl}
\langle V_{rh} \rangle\simeq \left(1+\frac{9}{2}\Delta^2 N_e\right) {e^{3 N_e} \over H^3(t_{hc})}\ .
\ee
The factor $e^{3 N_e}$ represents the classical expansion from the horizon crossing time to the time of reheating. The prefactor is a positive correction that is due to the short scale fluctuations that freeze from the time when the mode we consider crossed the horizon up to reheating.  These fluctuations enhance the overall expansion undergone by the universe in that amount of time. This can be thought of as a quantum generalization of the classical expansion history: the universe expands more that what would have done classically. This enhancement is proportional to the power spectrum of the fluctuations and to the classical number of $e$-foldings.

Fluctuations that crossed the horizon at a time $t_{hc}$ will appear on a scale containing a mass proportional to $V_{rh}$ in equation (\ref{eq:kinematical-one-taylor-Ncl}) and will have an amplitude given by equation (\ref{amplitude}). Note that the enclosed mass in completely independent of the late time history of the Universe, the rate of growth of structure, etc. If we wanted to compute a lengthscale for the mode, we can just take the cubic root of equation (\ref{eq:kinematical-one-taylor-Ncl}). 

Equation (\ref{eq:kinematical-one-taylor-Ncl}) implies that an observed scale crossed the horizon at a different point during the inflationary history. 
At lowest order in slow-roll it is easy to obtain an expression for the change in the value of $N_e$,
\be
\delta N_e = - {3 \over 2} \Delta^2 N_e\ . 
\ee
That is to say a fixed observed scale crossed the horizon closer to the end of inflation as a result of the enhanced expansion. This means that the amplitude of fluctuations on that scale is different. In fact it is equal to the amplitude of fluctuations that modes of a shorter scale would have in the absence of the effect we are considering.
% because these would be the modes that in the standard calculation would have crossed the horizon at that point in the inflationary history. 
As a result the amplitude of fluctuations due to our effect is directly given by the tilt:
\be\label{eq:deltaprojection}
\delta \ln (\Delta^2) = {3 \over 2} (n_s-1) \Delta^2 N_e\ .
\ee

We stress that this effect should not be interpreted as a time-dependence of $\zeta$. It is simply the projection effect of a given wavelength due to the subsequent expansion of the universe. Already at tree-level, that is neglecting short scale fluctuations, there are such effects because the final projected scale of a comoving wavenumber depends on the subsequent history of the universe. This happens for example through the number of inflationary $e$-foldings or through the reheating temperature. Eq.~(\ref{eq:deltaprojection}) is a quantum enhancement of the projection effect due to the short scale fluctuations that happens during the inflationary period. Short scale fluctuations affect the projection of a long mode even after the inflationary phase: as an example in~\cite{Baumann:2010tm,Carrasco:2012cv} it was shown that the presence of short scale inhomogeneities changes the equation of state $w$, and so the overall expansion, of a matter dominated universe with large scale gravitational potential fluctuations of order $10^{-5}$ from $w=0$ to~$w\sim 10^{-5}$.

So far we have approximated the stochastic enhancement of the expansion $e^{3(\zeta(t_{rh})-\zeta(t_{hc}))}$ with its expectation value. This is a good approximation only if its variance is small. This is indeed the case. In fact, the physical effect at the core of the enhanced expansion can be depicted in Fig.~\ref{Projection}. As the mode we consider crosses the horizon and continues to expand, modes of shorter wavelength keep coming out of the horizon and generate large fluctuations. Fluctuations will push the inflation forward or backward in its trajectory lengthening or shortening the duration of inflation in that patch. However, since backward fluctuations lead to more expansion and therefore to more fluctuations that can result in additional enhanced expansion, the overall effect is to enhance the expansion of the universe. This is the effect on average. 

\begin{figure}
\begin{center}
\includegraphics[width=16cm]{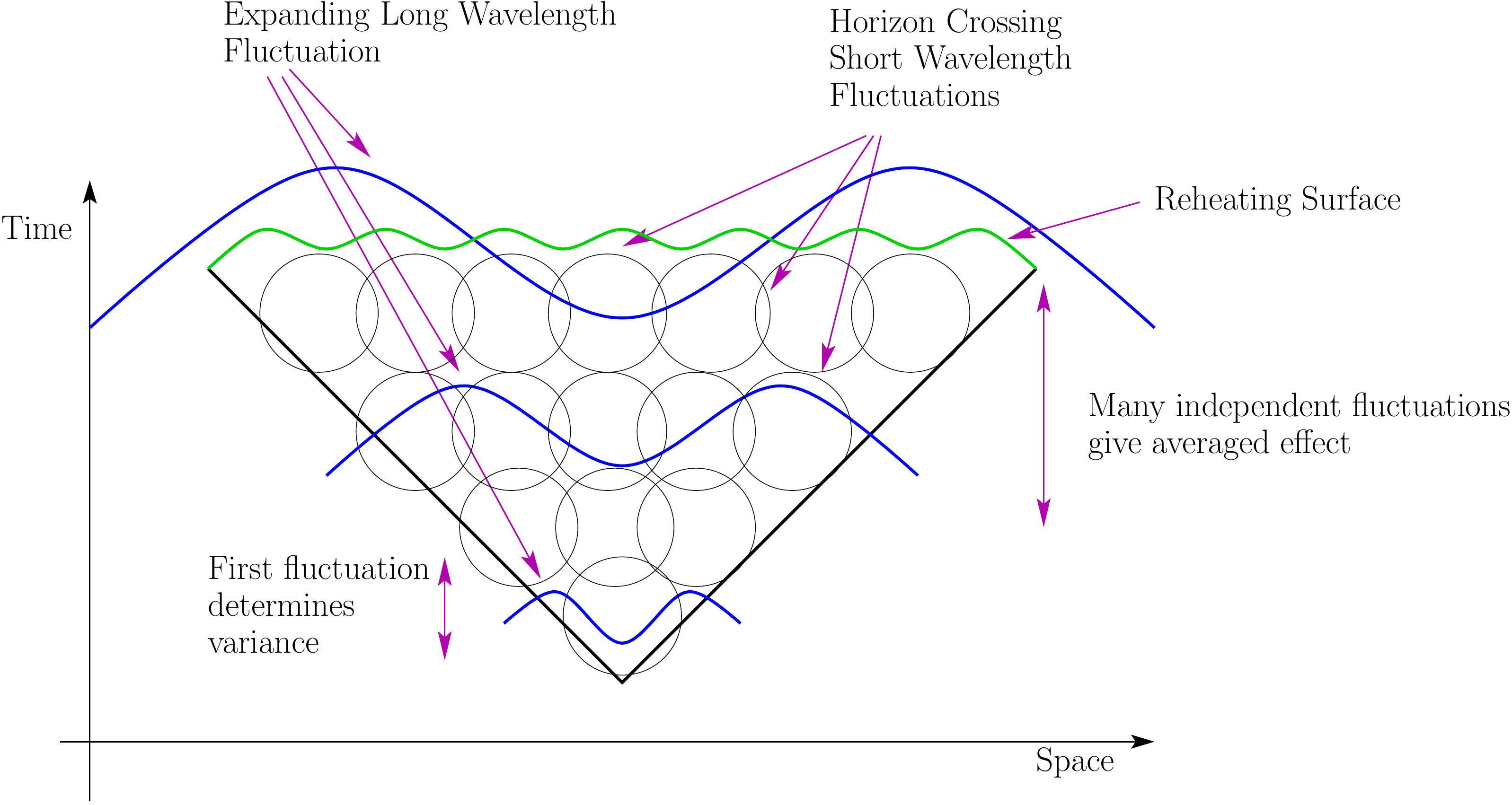}
\caption{\label{Projection} \small A long wavelength mode crosses the horizon and keeps expanding. Short wavelength modes continuously come out of the horizon. They affect the duration of inflation in the region, ultimately changing the physical volume $V_{rh}$ spanned by one comoving wavenumber $k$ on the reheating surface. The effects is to increase the average of this volume by an amount of order $\delta V_{rh}\sim V_{rh}\, \Delta \, N_e$. The factor of $N_e$ implies that the effect increases with the number of $e$-folding to the end of Inflation. The variance of this increase in volume is only proportional to $\Delta$, with no $N_e$ enhancement and it is determined by the fluctuations that cross the horizon within a few Hubble times after the mode $k$ crossed. The effect for subsequent fluctuations is very well approximated by its average, as there are many of these modes inside the mode $k$.}
\end{center}
\end{figure}

What about the variance? The variance results from the variance in the amplitudes of the fluctuations of all the short modes. Now, if the mode is much shorter than the one we consider there are many of those fluctuations within one wavelength of the mode of interest. Therefore their contribution to the variance will be small as there are many independent realizations inside the mode of interest.  However for the order one number of modes that come out of the horizon a few $e$-folding after our mode has crossed the horizon (see Fig.~\ref{Projection}), the variance from realization to realization will be significant.  Their contribution to the variance of $V_{rh}$ will simply be the variance of $\Delta^2$, which has no enhancement of $N_e$. A simple computation using equation (\ref{reheating-volume2}) agrees with this intuition. 
In summary the variance of $V_{rh}$ is not enhanced by $N_e$ and is simply of order
\be
\left({\rm Variance}\left[V_{rh}\right]\right)^{1/2} = \left(\langle V_{rh}^2 \rangle - \langle V_{rh}\rangle^2\right)^{1/2} \sim \Delta^2\ .
\ee
The variance is smaller than the change in expectation value by a factor of $N_e$. This is a nice result. The quantum-enhanced expansion is indeed of stochastic nature, but to a very good approximation has no variance: a mode that crossed the horizon at a given point during inflation will not end up having a scale given by the classical formula, but it will end up always basically on  the same scale in every realization.

Finally one could wonder about our choice of distance measure. Rather than using the volume of the enclosed region, one might want to use something else such as directly the distance between two points along a geodesic of the perturbed or unperturbed metrics. In general these different measures will be non-linearly related and one might wonder if one would get significantly different results if one were to study different cases. However, results will be the basically the same if one wants to track only terms enhanced by $N_e$. This is so because although the different quantities are stochastic, their variance is not enhanced. Thus when one tracks only terms enhanced by $N_e$ one can think of the mapping between horizon crossing and scale to be deterministic.
Take for example the difference between the distance along the perturbed or unperturbed geodesics. The correction from the standard answer for the two results will appear different, even the coefficient in front of the factor of $N_e$ will differ. However before interpreting these corrections one needs to relate them to distances as measured by a late time observer. Once this is done, the apparent difference will disappear. If for example one uses enclosed mass to measure distance as we have done in this paper, the apparent differences between the scaling of $N_e$ will disappear and will be compensated by a difference in the mapping between the two distances and enclosed mass which is non-linear but not stochastic. In this paper we chose the encoded mass because it is well defined, easy to compute and directly related to observable physical quantities~\footnote{In some naive sense, our prescription amounts to a rewriting of the usual power spectrum in different, more physical, length scales. This however is non-trivial, because in our variables effects of modes that are longer than the horizon trivially decouple from the calculation, so that it is made manifest that there are no physical IR divergencies coming from modes with longer wavelength on the surface of reheating. Notice additionally that this is a very unusual change of coordinates. Usually changes of coordinates are done on a manifold and are deterministic functions of the coordinates. Instead, the mapping from comoving coordinates to our physical variables is a function of a stochastic quantity $\zeta$, which changes on a realization by realization basis, and that only approximately can be replaced with its average. On every realization, not just on average, the effect of long modes cancels exactly once we go to the physical coordinates defined in that realization.}. 

Let us summarize what we found so far by saying that 
the physical scale on which modes that cross the horizon at a given time during inflation is projected, is in general different from what the standard one would give. This is due to the effect of modes shorter than the one we consider, modes that come out of the horizon after that mode. There is no effect from longer modes. There is thus a physical, potentially observable effect: a mode on a given physical scale has crossed the horizon at a different time than what would have been computed classically. This effect is enhanced by $N_e$, the number of $e$-foldings to the end of inflation. In our Universe this effect is of course extremely small.

\paragraph{Re-summing the kinematical effects and eternal inflation:}

Although not true in our observable universe, the expectation value of  $e^{3(\zeta(t_{rh})-\zeta(t_{hc}))}$ could becomes large if   $N_e\Delta^2 \gtrsim 1$. 
 In fact, this enhanced quantum expansion that we have been discussing is nothing but the perturbative version of slow-roll eternal inflation. Slow-roll eternal inflation is the limiting case in which the enhanced quantum expansion is so large to give a non-zero probability for inflation never to end. This suggests looking at the literature of slow eternal inflation.  Indeed, the probability distribution for the resulting volume as a function of the classical number of $e$-foldings and related quantities have been computed both in the regime of eternal and non-eternal inflation \cite{Creminelli:2008es,Dubovsky:2008rf,Dubovsky:2011uy, perko}. The analysis is quite complex so, we will not present it in detail here. We simply quote the final result for the expectation value and the variance of $V_{rh}$ in the regime far from eternal inflation. 
They read~\cite{Dubovsky:2008rf},
\bea
&&\langle e^{3(\zeta(t_{rh})-\zeta(t_{hc}))}\rangle\simeq {\rm Exp}\left[\frac{6N_e}{1+\sqrt{1-1/\Omega}}\right]\ , \qquad \Omega=-\frac{4\pi^2}{3}\frac{\dot H\mpl^2}{H^4}\ , \\ \nonumber
&&{\rm Variance}[e^{3(\zeta(t_{rh})-\zeta(t_{hc}))}]\simeq\frac{1}{\Omega}\left(1+\sqrt{1-\frac{1}{\Omega}}\right)^{-1}\ .
\eea
Eternal Inflation happens when $\Omega< 1$. Notice that if we expand perturbatively ($1/\Omega \ll 1$) the answer agrees we the one we had before.  We see that the enhanced overall expansion can be very large, but it is at most of order the double of the classical number of $e$-foldings, with corrections suppressed by slow-roll parameters, and it occurs at the transition to eternal inflation \cite{Dubovsky:2008rf}. The variance is always at most of order one number of $e$-folding, even in the regime close to eternal inflation.

\section{Conclusions}

Although coupling between modes of very different scales does not have any significant dynamical effect during inflation it can lead to interesting projection effects. These modes change the relation between scale and time of horizon crossing. We have argued that there are no infrared projection effects in physical questions, that is there are no effects from modes of longer wavelength than the one of interest. These potential effects cancel when computing fluctuations as a function of physically measurable scales. As an illustration we used the enclosed number of particles, or enclosed mass, as our measure of size. 
If we had computed the amplitude of fluctuations at a fixed comoving scale, we would have found an IR divergence. In our case this divergence is just a consequence of not having asked a physically relevant question. It can be made equal to zero in the extreme squeezed limit by a proper redefinition of the observables. 
%In some sense the statement is even stronger than in the standard examples of IR divergences, such as the soft photons in QED, where there is a finite physical effect. 
%In our present example, effects can be made exactly equal to zero by a proper redefinition of the observables. 

Modes on scales smaller than the one of interest do induce projection effects as defined in this paper. They change the mapping between observed scale and time of horizon crossing. The correction to the classical mapping is enhanced by a factor of $N_e$. Although this mapping is stochastic, its variance is not enhanced by $N_e$ so it is in a sense almost deterministic. There is a direct connection between the results we presented and studies of slow-roll eternal inflation. The full distribution for the volume at reheating has been computed in those studies. We made use of that literature to re-sum the perturbative calculation and get a result which is valid even if $N_e \Delta^2 \gg 1$.

If one is interested in predicting the outcome of a specific type of observation in the post-inflationary universe, the calculations in this paper are not a replacement for doing the full calculation. In general all modes inside the horizon at the time of observation will have both a dynamical and a projection effect on a mode of interest. The details will depend on the observable in question. As we stated before, in many circumstances these effects have been studied in the astrophysics literature. The effects we computed are small so studies have focused on other statistics such as the bispectrum, which in the case of the CMB is marginally detectable. A consistent calculation using the standard techniques of any measurable quantity would include the effect uncovered in this paper, but we have found it using rather simple physical arguments.

\subsubsection*{Acknowledgments}

We thank Nima Arkani-Hamed, Steve Giddings, Richard Holman, Shamit Kachru, Juan Maldacena, Steve Shenker, Eva Silverstein and Lenny Susskind for interesting conversations. L.S. is supported by the National Science Foundation under PHY-1068380.
M.Z. is supported by the National Science Foundation under PHY-
0855425 and AST-0907969 and by the David and Lucile 
Packard Foundation and the John D. and Catherine~T.~MacArthur~Foundation.

\appendix
\section*{Appendix}

In this appendix we make contact with existing calculations in the literature that compute the two point function at some given {\it comoving} momentum $k$~\cite{Giddings:2010nc,Byrnes:2010yc}. 
At issue are the effects of the long wavelength modes and the associated infrared divergencies. In any realization, even before averaging over the long modes, the amplitude of fluctuations at a smaller scale $k$ is given by:
\be\label{app1}
\Delta^2_B(k) = \Delta^2_0(k e^{-\zeta_B})\ .
\ee
All we need to do is expand equation (\ref{app1}) in Taylor series,
\be
\Delta^2_B(k) \approx \Delta^2_0- {\partial \Delta^2_0 \over \partial \ln k} \zeta_B +{1\over 2}  {\partial^2 \Delta^2_0 \over \partial \ln k^2} \zeta_B^2 + \cdots\ ,
\ee
and take expectation value over the long modes:
\be
\langle\Delta^2_B\rangle \approx \Delta^2_0 +{1\over 2}  {\partial^2 \Delta^2_0 \over \partial \ln k^2} \langle \zeta_B^2 \rangle+ \cdots\ .
\ee
We can use the standard definitions of the tilt ($n_s$) and the running ($\alpha$):
\be
n_s-1 = {\partial \ln \Delta^2_0 \over \partial \ln k} \ \ \ \ ; \ \ \ \  \alpha = {\partial n_s \over \partial \ln k}\ ,
\ee
to obtain:
\be
\langle\Delta^2_B\rangle \approx \Delta^2_0(k) \left[1+ {1\over 2} \left((n_s-1)^2+\alpha\right) \langle \zeta_B^2 \rangle+ \cdots\right]\ .
\ee
Now the expectation value is:
\be\label{eq:N-beginning}
\langle\zeta_B^2\rangle=\int_{k^\prime < k} d\ln k^\prime \Delta^2(k^\prime)\approx\Delta_0 \log(k L)\simeq \Delta_0 N_e^{\rm beginning} \ ,
\ee
where ${k^\prime < k}$ characterizes the modes that we are including in the background for a given $k$ and $\log(k L)\simeq  N_e^{\rm beginning}$ is the number of $e$-foldings from the beginning of inflation to when the mode $k$ crosses the horizon and $L$ is an IR cut-off. Notice that $N_e^{\rm beginning}$ can be a huge number, possibly even without an upper bound as in eternal inflation, where $\Delta_0\sim 1$, further enhancing the effect. Although this result might seem naively worrisome it is not. We have already pointed out that this result is equivalent to the statement that long wavelength modes have no physical effect. Thus this apparent divergence disappears in a calculation of a physical observable.

 \begingroup\raggedright\endgroup

%\bibliography{trapbib}

\begin{thebibliography}{10}

%\cite{Senatore:2009cf}
\bibitem{Senatore:2009cf}
  L.~Senatore and M.~Zaldarriaga,
  ``On Loops in Inflation,''
  JHEP {\bf 1012} (2010) 008
  [arXiv:0912.2734 [hep-th]].
  %%CITATION = JHEPA,1012,008;%%

%\cite{Weinberg:2005vy}
\bibitem{Weinberg:2005vy}
  S.~Weinberg,
  ``Quantum contributions to cosmological correlations,''
  Phys.\ Rev.\  D {\bf 72} (2005) 043514
  [arXiv:hep-th/0506236].
  %%CITATION = PHRVA,D72,043514;%%

%\cite{Weinberg:2006ac}
\bibitem{Weinberg:2006ac}
  S.~Weinberg,
  ``Quantum contributions to cosmological correlations. II: Can these
  corrections become large?,''
  Phys.\ Rev.\  D {\bf 74} (2006) 023508
  [arXiv:hep-th/0605244].
  %%CITATION = PHRVA,D74,023508;%%



%\cite{Kahya:2010xh}
\bibitem{Kahya:2010xh}
  E.~O.~Kahya, V.~K.~Onemli and R.~P.~Woodard,
  ``The Zeta-Zeta Correlator Is Time Dependent,''
  Phys.\ Lett.\  B {\bf 694} (2010) 101
  [arXiv:1006.3999 [astro-ph.CO]].
  %%CITATION = PHLTA,B694,101;%%


%\cite{Giddings:2010nc}
\bibitem{Giddings:2010nc}
  S.~B.~Giddings and M.~S.~Sloth,
  ``Semiclassical relations and IR effects in de Sitter and slow-roll
  space-times,''
  JCAP {\bf 1101} (2011) 023
  [arXiv:1005.1056 [hep-th]].
  %%CITATION = JCAPA,1101,023;%%

 
%\cite{Byrnes:2010yc}
\bibitem{Byrnes:2010yc} 
  C.~T.~Byrnes, M.~Gerstenlauer, A.~Hebecker, S.~Nurmi and G.~Tasinato,
  ``Inflationary Infrared Divergences: Geometry of the Reheating Surface versus $\delta N$ Formalism,''
  JCAP {\bf 1008}, 006 (2010)
  [arXiv:1005.3307 [hep-th]].
  %%CITATION = ARXIV:1005.3307;%%   

%\cite{Gerstenlauer:2011ti}
\bibitem{Gerstenlauer:2011ti}
  M.~Gerstenlauer, A.~Hebecker and G.~Tasinato,
  ``Inflationary Correlation Functions without Infrared Divergences,''
  JCAP {\bf 1106} (2011) 021
  [arXiv:1102.0560 [astro-ph.CO]].
  %%CITATION = JCAPA,1106,021;%%


%\cite{Giddings:2011zd}
\bibitem{Giddings:2011zd}
  S.~B.~Giddings and M.~S.~Sloth,
  ``Cosmological observables, IR growth of fluctuations, and scale-dependent
  anisotropies,''
  Phys.\ Rev.\  D {\bf 84} (2011) 063528
  [arXiv:1104.0002 [hep-th]].
  %%CITATION = PHRVA,D84,063528;%%


%\cite{Giddings:2011ze}
\bibitem{Giddings:2011ze} 
  S.~B.~Giddings and M.~S.~Sloth,
  ``Fluctuating geometries, q-observables, and infrared growth in inflationary spacetimes,''
  arXiv:1109.1000 [hep-th].
  %%CITATION = ARXIV:1109.1000;%%


   
%\cite{loops1}
\bibitem{loops1}
 G.~Pimentel, L.~Senatore, M.~Zaldarriaga
  ``On Loops in Inflation III: time-independence of $\zeta$ correlators'', to appear.

%\cite{Chialva:2011bg}
\bibitem{Chialva:2011bg}
  D.~Chialva and A.~Mazumdar,
  ``Eliminating infrared divergences in an inflationary cosmology,''
  arXiv:1103.1312 [hep-th].
  %%CITATION = ARXIV:1103.1312;%%

%\cite{Tanaka:2011aj}
\bibitem{Tanaka:2011aj}
  T.~Tanaka and Y.~Urakawa,
  ``Dominance of gauge artifact in the consistency relation for the primordial bispectrum,''
  JCAP {\bf 1105} (2011) 014
  [arXiv:1103.1251 [astro-ph.CO]].
  %%CITATION = ARXIV:1103.1251;%%

 
 %\cite{Urakawa:2010kr}
\bibitem{Urakawa:2010kr}
  Y.~Urakawa and T.~Tanaka,
  ``Natural selection of inflationary vacuum required by infra-red regularity
  and gauge-invariance,''
  Prog.\ Theor.\ Phys.\  {\bf 125} (2011) 1067
  [arXiv:1009.2947 [hep-th]].
  %%CITATION = PTPKA,125,1067;%%



%\cite{Urakawa:2011fg}
\bibitem{Urakawa:2011fg}
  Y.~Urakawa,
  ``Influence of gauge artifact on adiabatic and entropy perturbations during inflation,''
  Prog.\ Theor.\ Phys.\  {\bf 126} (2011) 961
  [arXiv:1105.1078 [hep-th]].
  %%CITATION = ARXIV:1105.1078;%%


%\cite{Riotto:2011sf}
\bibitem{Riotto:2011sf} 
  A.~Riotto and M.~S.~Sloth,
  ``The probability equation for the cosmological comoving curvature perturbation,''
  JCAP {\bf 1110}, 003 (2011)
  [arXiv:1103.5876 [astro-ph.CO]].
  %%CITATION = ARXIV:1103.5876;%%











 
%\cite{Ganc:2010ff}
\bibitem{Ganc:2010ff}
  J.~Ganc and E.~Komatsu,
  ``A new method for calculating the primordial bispectrum in the squeeze limit,''
  JCAP {\bf 1012} (2010) 009
  [arXiv:1006.5457 [astro-ph.CO]].
  %%CITATION = JCAPA,1012,009;%%

%\cite{Creminelli:2011rh}
\bibitem{Creminelli:2011rh}
  P.~Creminelli, G.~D'Amico, M.~Musso and J.~Norena,
  ``The (not so) squeezed limit of the primordial 3-point function,''
  JCAP {\bf 1111} (2011) 038
  [arXiv:1106.1462 [astro-ph.CO]].
  %%CITATION = JCAPA,1111,038;%%


%\cite{Cheung:2007st}
\bibitem{Cheung:2007st}
  C.~Cheung, P.~Creminelli, A.~L.~Fitzpatrick, J.~Kaplan and L.~Senatore,
  ``The Effective Field Theory of Inflation,''
  JHEP {\bf 0803} (2008) 014
  [arXiv:0709.0293 [hep-th]].
  %%CITATION = JHEPA,0803,014;%%

%\cite{Nacir:2011kk}
\bibitem{Nacir:2011kk}
  D.~L.~Nacir, R.~A.~Porto, L.~Senatore and M.~Zaldarriaga,
  ``Dissipative effects in the Effective Field Theory of Inflation,''
  JHEP {\bf 1201} (2012) 075
  [arXiv:1109.4192 [hep-th]];
  %%CITATION = JHEPA,1201,075;%%
   %\cite{Green:2009ds}
   %   \bibitem{Green:2009ds}
        D.~Green, B.~Horn, L.~Senatore and E.~Silverstein,
        ``Trapped Inflation,''
        Phys.\ Rev.\  D {\bf 80} (2009) 063533
        [arXiv:0902.1006 [hep-th]];
        %%CITATION = PHRVA,D80,063533;%%
%\cite{Senatore:2011sp}
%\bibitem{Senatore:2011sp}
  L.~Senatore, E.~Silverstein and M.~Zaldarriaga,
  ``New Sources of Gravitational Waves during Inflation,''
  arXiv:1109.0542 [hep-th].
  %%CITATION = ARXIV:1109.0542;%%

%\cite{Baumann:2010tm}
\bibitem{Baumann:2010tm}
  D.~Baumann, A.~Nicolis, L.~Senatore and M.~Zaldarriaga,
  ``Cosmological Non-Linearities as an Effective Fluid,''
  arXiv:1004.2488 [astro-ph.CO].
  %%CITATION = ARXIV:1004.2488;%%

%\cite{Carrasco:2012cv}
\bibitem{Carrasco:2012cv}
  J.~J.~M.~Carrasco, M.~P.~Hertzberg and L.~Senatore,
  ``The Effective Field Theory of Cosmological Large Scale Structures,''
  arXiv:1206.2926 [astro-ph.CO].
  %%CITATION = ARXIV:1206.2926;%%

%\cite{Creminelli:2008es}
\bibitem{Creminelli:2008es} 
  P.~Creminelli, S.~Dubovsky, A.~Nicolis, L.~Senatore and M.~Zaldarriaga,
  ``The Phase Transition to Slow-roll Eternal Inflation,''
  JHEP {\bf 0809}, 036 (2008)
  [arXiv:0802.1067 [hep-th]].
  %%CITATION = ARXIV:0802.1067;%%
  
 %\cite{Dubovsky:2008rf}
\bibitem{Dubovsky:2008rf} 
  S.~Dubovsky, L.~Senatore and G.~Villadoro,
  ``The Volume of the Universe after Inflation and de Sitter Entropy,''
  JHEP {\bf 0904}, 118 (2009)
  [arXiv:0812.2246 [hep-th]].
  %%CITATION = ARXIV:0812.2246;%%
  
 %\cite{Dubovsky:2011uy}
\bibitem{Dubovsky:2011uy} 
  S.~Dubovsky, L.~Senatore and G.~Villadoro,
  ``Universality of the Volume Bound in Slow-Roll Eternal Inflation,''
  arXiv:1111.1725 [hep-th].
  %%CITATION = ARXIV:1111.1725;%%
   

  %\cite{perko}
\bibitem{perko}
  M. Lewandowski, A.~Perko, L.~Senatore and G.~Villadoro,
  ``The Volume of the Universe after Slow-Roll Corrections  and its Universal Bound,''
 in preparation. 





\end{thebibliography}
\end{document}